\documentclass[12pt]{article}
\usepackage{graphicx}

\newsavebox{\sboxpubnumber}
\newsavebox{\sboxpubdate}
\newcommand{\pubdate}[1]{\begin{lrbox}{\sboxpubdate}{#1}\end{lrbox}}
\newcommand{\pubnumber}[1]{\begin{lrbox}{\sboxpubnumber}{\begin{tabular}{l} #1 \\
				 \usebox{\sboxpubdate}
				 \end{tabular}}
                           \end{lrbox}
                           \pubblock}
\newcommand{\Title}[1]{\begin{center} {\Large #1 } \end{center}}
\newcommand{\Author}[1]{\begin{center}{ \sc #1} \end{center}}
\newcommand{\Address}[1]{\begin{center}{ \it #1} \end{center}}

\newcommand{\pubblock}{\rightline{
			\usebox{\sboxpubnumber}}}
\newenvironment{Abstract}{\begin{quotation}  }{\end{quotation}}
\newenvironment{Presented}{\begin{quotation} \begin{center}
             PRESENTED AT\end{center}\bigskip
      \begin{center}\begin{large}}{\end{large}\end{center}
      \end{quotation}}
\newcommand{\Acknowledgements}{\bigskip  \bigskip \begin{center} \begin{large}
             \bf ACKNOWLEDGEMENTS \end{large}\end{center}}
\begin{document}

%%%%%%%%%%%%%%%%%%%%%%%%%%%%%%%%%%%%%%%%%%%%%%%%%%%%%%%%%%%%%%%%%%%%%%%%
%%
%% START EDITING HERE!
%%
%%%%%%%%%%%%%%%%%%%%%%%%%%%%%%%%%%%%%%%%%%%%%%%%%%%%%%%%%%%%%%%%%%%%%%%%
\begin{titlepage}
\pubdate{\today}                    %fill in the date
\pubnumber{NORDITA-2001-96-AP} %preprint number(s)

\vfill
\Title{Recent Progress in Neutrino Astrophysics}
\vfill
\Author{Steen Hannestad}
\Address{NORDITA \\
         Blegdamsvej 17, DK-2100 Copenhagen, Denmark}
\vfill
\begin{Abstract}
Recent progress in neutrino physics has been rapid, to a large
extent thanks to observations of neutrinos produced in astrophysical
environments. Here, we review the current standing on such questions
as neutrino masses and mixings, focusing mainly on the interplay
between neutrino physics, astrophysics and cosmology.
\end{Abstract}
\vfill
\begin{Presented}
    COSMO-01 \\
    Rovaniemi, Finland, \\
    August 29 -- September 4, 2001
\end{Presented}
\vfill
\end{titlepage}
\def\thefootnote{\fnsymbol{footnote}}
\setcounter{footnote}{0}

%%%%%%%%%%%%%%%%%%%%%%%%%%%%%%%%%%%%%%%%%%%%%%%%%%%%%%%%%%%%%%%%%%%%%%%%
% The document starts here
%%%%%%%%%%%%%%%%%%%%%%%%%%%%%%%%%%%%%%%%%%%%%%%%%%%%%%%%%%%%%%%%%%%%%%%%
\section{Introduction}

The field of neutrino physics is presently advancing more rapidly than
ever before. For a large part this is due to observations of neutrinos
coming from astrophysical sources, such as the Sun, cosmic rays and
supernovae.
Neutrino astrophysics has therefore been a key ingredient in retrieving
the fundamental parameters related to the neutrino sector of the
standard model.
From a purely phenomenological perspective (as opposed to the 
model building perspective), the neutrino sector can be described
by a relatively limited number of parameters:
The neutrino masses, the neutrino mixing matrix which is the equivalent
of the quark CKM matrix, as well as the possible cosmological neutrino
lepton number.
In addition there is the possibility of parameters describing neutrino
physics beyond the standard model: Sterile neutrino masses and mixing,
neutrino magnetic moments, flavour changing neutral currents etc.
The next section contains a review of the current knowledge about
neutrino masses differences and mixings, obtained from neutrino oscillation 
experiments.
Section 3 deals with neutrinos and cosmology, particularly the
prospects for measuring absolute neutrino masses and cosmological
neutrino lepton numbers.
Section 4 is on supernova neutrinos, and finally Section 5 contains 
a short discussion.

%%%%%%%%%%%%%%%%%%%%%%%%%%%%%%%%%%%%%%%%%%%%%%%%%%%%%%%%%%%%%%%%%%%%%%%%
% Section II
%%%%%%%%%%%%%%%%%%%%%%%%%%%%%%%%%%%%%%%%%%%%%%%%%%%%%%%%%%%%%%%%%%%%%%%%

\section{Neutrino oscillations}

\subsection{Solar neutrinos}

The Sun is by far the most powerful nearby neutrino source. 
The hydrogen fusion reactions which heat the Sun also produce large
numbers of neutrinos via the effective reaction
$4{\rm H} + 2e^- \to ^4{\rm He} + 2 \nu_e$. For low mass stars like the Sun
almost all energy is produced via the pp-chain, as opposed to
more massive stars where the CNO cycle is dominant.

In fact neutrinos are produced via several different nuclear reactions.
The bulk of all neutrinos are so-called pp-neutrinos produced in
the reaction $p+p \to ^2 {\rm H} + e^+ + \nu_e$. However, these neutrinos
are born with very low energy ($E \leq 0.42$ MeV) and are therefore
very difficult to detect. For detection experiments the 
most important reactions are $e^- + ^7 {\rm Be} \to ^7 {\rm Li} + \nu_e$ 
(Beryllium neutrinos, $E_\nu = 0.86$ MeV) and
$p + ^7 {\rm Be} \to ^8 {\rm B} \to ^8 
{\rm Be} + e^+ + \nu_e$ (Boron neutrinos,
$E \leq 15$ MeV). Figure 1 shows the Solar neutrino flux on
earth according to the latest standard solar model \cite{bp00}.
%%%%%%%%%%%%%%%%%%%%%%%%%%%%%%%%%%%%%%%%%%%%%%%%%%%%%%%%%%%%%%%%%%%%%%%%
%%
%%   use this format to include an .eps figure into your paper
%%
\begin{figure}[htb]
    \centering
    \rotatebox{-90}{\includegraphics[height=5.5in]{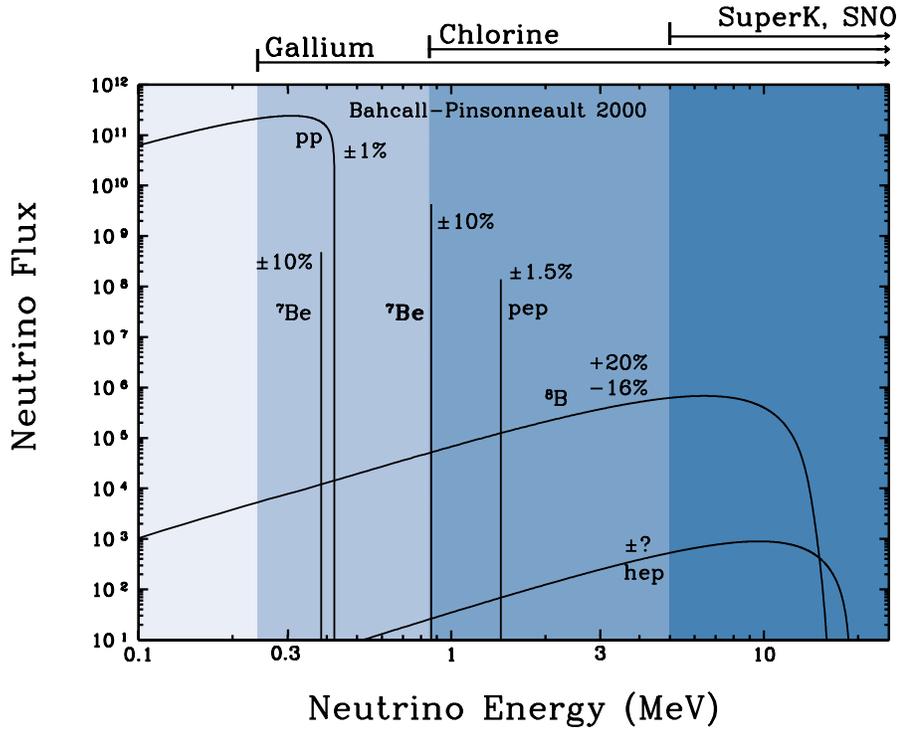}}
    \caption{Neutrino flux prediction from the present standard
solar model \cite{bp00}. Figure courtesy of J.~Bahcall.}
    \label{fig:bahcall2}
\end{figure}
%%%%%%%%%%%%%%%%%%%%%%%%%%%%%%%%%%%%%%%%%%%%%%%%%%%%%%%%%%%%%%%%%%%%%%%%

Although solar
neutrinos are of low energy and therefore have very small scattering
cross section they are detectable in terrestrial detectors. 
The first such experiment was the Homestake experiment 
\cite{Lande:cv,davis} which used a
radiochemical observation of conversion of $^{37}$Cl into $^{37}$Ar 
for detection.

The experiment first published data in 1968 \cite{davis}, and at 
the same time more detailed Solar models
became available \cite{solar}. 
It then became evident that there was a significant 
discrepancy between theory and observations.
This discrepancy has persisted more or less unchanged until now, and
is seen for all types of experiments. There are basically two different
techniques for detecting solar neutrinos. One is the radiochemical 
method of which the Homestake experiment was the first example.
Later experiments have used Ga instead of Cl because a much lower
energy threshold is possible. Indeed, the SAGE \cite{Abdurashitov:cw}
and Gallex
\cite{Hampel:1998xg,Cribier:am} experiments
have been able to detect the low energy pp-neutrinos.

The second is the water Cerenkov detector, where the reaction
$\nu_e + e \to \nu_e + e$ is used. For MeV neutrinos scattering on
electrons, the final state energy of the electron is sufficiently 
high that it emits Cerenkov light as it travels through water.
This means that both energy and direction of the incoming neutrino
can be measured. However, the energy threshold of Cerenkov detectors
is $\sim$5-8 MeV. Therefore only the high energy tail of the solar
neutrino distribution, mainly the $^8$B neutrinos are detectable.
Radiochemical experiments have much lower threshold, particularly
the gallium based experiments, and are therefore sensitive also 
to the main component of the the solar neutrino flux, the pp neutrinos.
The two types of experiments therefore complement each other nicely.

The fact that the solar neutrino flux is consistently about a
factor of two smaller than the theoretically expected value has been
seen as evidence for neutrino oscillations. However, before this
year there was no real smoking gun signature, because 
all that could be seen was disappearance of electron neutrinos.
In principle this could be ascribed to a lower primary flux of
electron neutrinos coming from the sun.

The new results from the Sudbury Neutrino Observatory (SNO)
\cite{Ahmad:2001an}, 
on the other hand, clearly show that there is an active component
in the solar neutrino flux which is not $\nu_e$. 
The reason is that SNO uses heavy water D$^2$O for detection instead
of ordinary water. This has the advantage that both the
charged current reaction $\nu_e + {\rm D} \to p + p + e^-$ and
the neutral current elastic scattering $\nu + e \to \nu + e$
can be detected, whereas normal water Cerenkov detectors are
only sensitive to the neutral current reaction.
By comparing the inferred neutrino fluxed from CC and NC reactions
it has been possible to ascertain (at the $3.3\sigma$ level) that
the NC measured flux is higher than the CC flux. This in turn
can only mean that there is an active component in the solar neutrino
flux which is not electron flavour.
Since the Sun
is not nearly hot enough to produce muon and tau neutrinos this component
can only have been produced via conversion of electron neutrinos.

This result also shows that the main part of the conversion of Solar 
neutrinos cannot be to sterile neutrinos, but must be to some active
species.
The combination of all available solar neutrino data still leaves
four distinct possible solar neutrino solutions. Table I shows
that best fit values from Ref.\ \cite{Bahcall:2001zu}
for these solutions, as well as their respective
goodness-of-fit (see also the analyses 
\cite{Krastev:2001tv,Berezinsky:2001se,Bandyopadhyay:2001aa}). 
Interestingly the so-called Small Mixing Angle solution
is now disfavoured, primarily because of the flatness of the energy
spectrum measured by Super Kamiokande \cite{Fukuda:2001nk}
and SNO \cite{Ahmad:2001an}.

\begin{table}
\caption{Best fit values of mixing parameters for solar neutrino
solutions, as well as their goodness of fit.
Taken from \cite{Bahcall:2001zu}.}
\begin{center}
\begin{tabular}{lccc} \hline \hline
Solution & $\delta m^2/{\rm eV}^2$ & $\tan ^ 2 \theta$ & goodness of fit \\ \hline 
Large Mixing Angle (LMA) & $4.5 \times 10^{-5}$ & 0.41 & 59\% \\
Small Mixing Angle (SMA) & $4.5 \times 10^{-5}$ & $3.9 \times 10^{-4}$ & 19\% \\
Low & $1.0 \times 10^{-7}$ & 0.71 & 45\% \\
Vacuum (VAC) & $4.6 \times 10^{-10}$ & 2.4 & 42\% \\ \hline \hline
\end{tabular}
\end{center}
\end{table}

Figure 2 shows a contour plot for the likelihood as a function of
$\tan^2 \theta$ and $\delta m^2$.
Notice that this figure is for the case where the $^8$B flux is taken
as a free parameter. However, the small mixing angle solution
prefers a value of the flux different from what is predicted by the
standard solar model. If the $^8$B flux is fixed by the SSM prediction
the exclusion of the SMA solution becomes much stronger
\cite{Bahcall:2001cb,Aliani:2001zi}. However,
as long as the spectrum information is not included the SMA solution
is actually preferred and perhaps it is premature to rule it out
completely. Fortunately the KamLand experiment 
will probe the LMA region of parameter space and
therefore test whether the LMA solution is indeed the correct one
\cite{Piepke:tg}.

%%%%%%%%%%%%%%%%%%%%%%%%%%%%%%%%%%%%%%%%%%%%%%%%%%%%%%%%%%%%%%%%%%%%%%%%
%%
%%   use this format to include an .eps figure into your paper
%%
\begin{figure}[htb]
    \centering
    \includegraphics[height=3.5in]{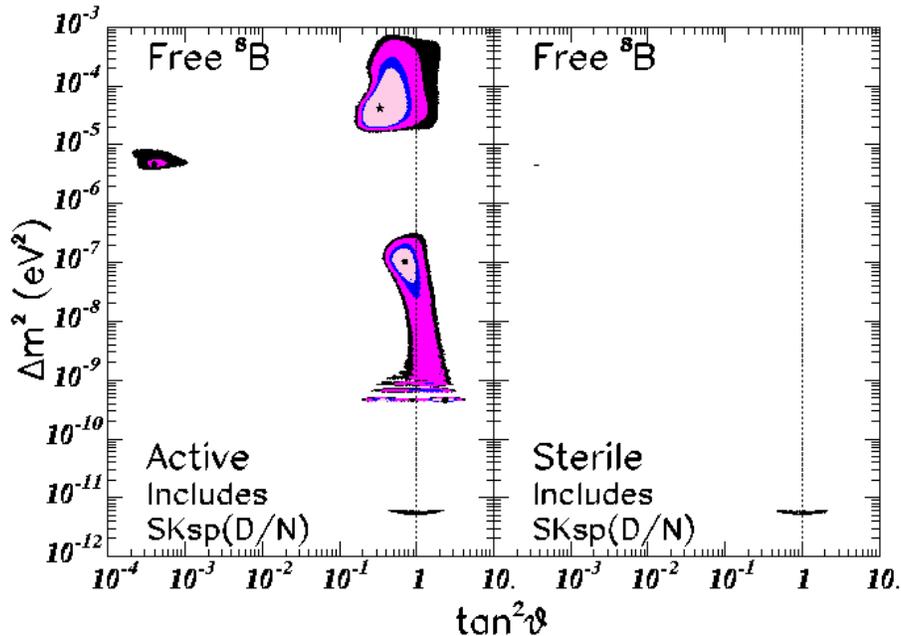}
    \caption{Likelihood plot including all available data on solar
neutrinos. Figure adapted from Ref.~\cite{Bahcall:2001zu}.}
    \label{fig:bahcall}
\end{figure}
%%%%%%%%%%%%%%%%%%%%%%%%%%%%%%%%%%%%%%%%%%%%%%%%%%%%%%%%%%%%%%%%%%%%%%%%

\subsection{Atmospheric neutrinos}

The second piece of evidence for neutrino oscillation comes from
observations of atmospheric neutrinos by Super Kamiokande \cite{Fukuda:1998mi}
and other neutrino detectors.
These neutrinos are produced by pion decays in the atmosphere, and 
therefore it is expected that the ratio of muon to electron
neutrinos should be roughly 2:1 (from the decay chain 
$\pi \to \mu \nu_\mu \to e \nu_e \nu_\mu \nu_\mu$). However, observations
clearly show that the ratio is close to 1:1. Again, this is 
explained by muon neutrinos oscillating with another species, with
close to maximal mixing. The real smoking gun signature for neutrino
oscillations is the so-called up-down asymmetry. For a given
solid angle the neutrino event rate should be independent of direction.
For electron neutrinos this is indeed the case, but for muon neutrinos
there is a much larger downwards flux. This can be explained by 
oscillations: Since upgoing neutrinos traverse a much longer distance
they have time to oscillate, whereas downgoing neutrinos do not.

Since no asymmetry is seen for electron neutrinos, the result cannot
be explained by $\nu_e - \nu_\mu$ oscillations, but is more
likely either $\nu_\mu - \nu_\tau$ or $\nu_\mu - \nu_s$ oscillations,
where $\nu_s$ is some sterile state.
Interestingly, the possibility that the oscillations are to a sterile
species is now also disfavoured because of the lack of a 
sterile neutrino matter potential in the earth. 
In Ref.\ \cite{Fornengo:2000sr} it was found that $\nu_\mu - \nu_\tau$ with 
$\delta m_{23}^2 = 3.0 \times 10^{-3}$ eV$^2$ and $\sin ^2 2 \theta = 0.99$
gives $\chi^2/{\rm d.o.f.} = 74/73$, whereas the
$\nu_\mu - \nu_s$ possibility
gives $\chi^2/{\rm d.o.f.} = 90/73 (\delta m^2 >0)
86/73 (\delta m^2 < 0)$.

A clear picture therefore seems to be emerging on neutrino 
mass differences. $\nu_e -\mu_\mu$ oscillations are responsible
for the Solar neutrino anomaly, whereas $\nu_\mu -\mu_\tau$ oscillations
explain the solar neutrino oscillations. From atmospheric neutrinos
one finds that $\delta m_{23}^2 \sim 3 \times 10^{-3}$ eV$^2$,
and $\sin ^2 2 \theta_{23} \simeq 1$. Solar neutrinos then 
give that $\delta m_{23}^2 = \delta m^2_{\rm solar}$ and
$\theta_{12} = \theta_{\rm solar}$.

The $1-3$ mixing angle has been constrained to be
rather small by the Chooz reactor experiment
\cite{Apollonio:1999ae}, and it therefore seems that
the neutrino mixings can be described effectively by $2\times2$ mixing.

However, the neutrino oscillations experiments do not yield any
information about the absolute values of the neutrino masses,
and therefore there are still two distinct possibilities, either
that the heaviest neutrino has a mass of 0.1 eV and that the 
lightest neutrino is close to massless.
The other possibility is a degenerate neutrino mass spectrum, where
all neutrinos have essentially the same mass, but with mass splitting
compatible with the solar and atmospheric neutrino solutions.
Cosmology is one of the strongest tools for probing the absolute 
values of neutrino masses, as will be discussed in the next section.

It should also be noted that there is at present no measurement
of the possible $CP$-violating phases in the neutrino mixing 
matrix. Therefore the work on measuring the mixing matrix entries
is far from finished.

%%%%%%%%%%%%%%%%%%%%%%%%%%%%%%%%%%%%%%%%%%%%%%%%%%%%%%%%%%%%%%%%%%%%%%%%
% Section III
%%%%%%%%%%%%%%%%%%%%%%%%%%%%%%%%%%%%%%%%%%%%%%%%%%%%%%%%%%%%%%%%%%%%%%%%

\section{Neutrinos in cosmology}

\subsection{Absolute neutrino masses}

Neutrinos exist in equilibrium with the electromagnetic plasma
in the early universe, until a temperature of a few MeV. At this
point the weak interactions freeze out and neutrinos decouple from
the plasma. Shortly after this time, the temperature of the plasma
falls below the electron mass, and electrons and positrons annihilate,
dumping their entropy into the photon gas. This heats the photon
gas while having no effect on neutrinos, and the end result is
that the photon temperature is larger than the neutrino temperature
by the factor $(11/4)^{1/3} \simeq 1.40$. Since the present day
photon temperature has been measured with great accuracy to be
2.728 K, the neutrino temperature is known to be 1.95 K, or
about $2 \times 10^{-4}$ eV. Since the heaviest neutrino has
a mass of at least about 0.1 eV it must at present be extremely
non-relativistic and therefore acts as dark matter.
The contribution of a single neutrino species of mass $m_\nu$
to the present day matter density can be written as
\cite{Kolb:vq,Cowsik:gh}
\begin{equation}
\Omega_\nu h^2 = \frac{m_\nu}{91.5 {\rm eV}},
\end{equation}
so that for a neutrino mass of about 30 eV, neutrinos will make
up all of the dark matter.
However, this would have disastrous consequences for structure
formation in the universe, because neutrinos of eV mass have very
large free streaming lengths and would erase structure on 
scales smaller than $l_{\rm fs} \simeq 1 \, {\rm Gpc} 
m_{\nu,{\rm eV}}^{-1}$
completely.
The linear matter power spectrum has been measured reasonably accurately
by large scale structure surveys down to scales of about 
$k \simeq 0.2 h {\rm Mpc}^{-1}$ (see for instance Ref.\ \cite{htp}).
In addition to this, recent very accurate measurements of 
the absorption power spectra of the Ly-$\alpha$ forest at high
redshift have allowed determination of the linear power spectrum
to even smaller scales \cite{croft}.
These power spectra can be used to constrain possible values of
$m_\nu$, particularly of used in combination with measurements
of the CMBR.
The best present upper limit on the neutrino mass is 4.4 eV \cite{wang},
which can be considered as a limit on the sum of masses of all the 
light neutrinos.

This upper limit can be compared with the present bound on the
electron neutrino mass from tritium decay endpoint measurements.
The Mainz experiment currently yields an upper limit of 2.2 eV
for the electron neutrino \cite{Bonn:tw}. 
If the mass is close to this bound the
neutrino mass hierarchy must be degenerate and the bound on the
sum of masses is therefore something like 6.6 eV.

However, it should be noted that an even stronger upper bound
can be put on neutrino masses if they are Majorana particles.
In that case neutrinoless double beta decay is possible
because lepton number is not a conserved quantity. The
non-observation of such events has led to the bound
\begin{equation}
m_{ee} = \sum_j U^2_{ej} m_{\nu_j} < 0.27 \,\, {\rm eV},
\end{equation}
where $U$ is the neutrino mixing matrix \cite{klapdor}.

In the coming years the large scale structure power spectrum will
be measured even more accurately by the Sloan Digital Sky Survey,
and at the same time the CMBR anisotropy will be probed to great
precision by the MAP and Planck satellites. By combining these
measurements it was estimated by Hu, Eisenstein
and Tegmark that a sensitivity of about 0.3 
eV could be reached \cite{Hu:1997mj}.
Currently an upgrade of the Mainz experiment by an order of magnitude
is planned, which should take the limit on the electron neutrino
mass down to about 0.2 eV.
The prospects for measuring a neutrino mass of the order 0.1 eV,
as suggested by oscillation experiments is therefore almost within reach.

Another cosmological probe of the neutrino mass is the so-called
Z-burst scenario for ultrahigh energy cosmic rays 
\cite{Weiler:1997sh,Fargion:1997ft}. 
Neutrinos are
not subject to the GZK cut-off which applies to protons
\cite{Greisen:1966jv,Zatsepin:1966jv}. Therefore it is
in principle possible that the primary particles for super GZK
cosmic rays are neutrinos. One possibility which has been explored
is that the neutrino-nucleon cross-section increases drastically
at high CM energies, for instance due to the presence of large 
extra dimensions.
The other possibility is that neutrinos have rest mass in the 
eV range. In that case high energy neutrinos can annihilate
on cosmic background neutrinos with a large cross section if the 
CM energy is close to the Z-resonance, corresponding to a primary
neutrino energy of $E_\nu \simeq 4 \times 10^{21} m_{\rm eV}^{-1} \, {\rm eV}$.
This annihilation would produce high energy protons which could then
act as primaries for the observed high energy cosmic rays.
The observed ultrahigh energy cosmic ray flux can be explained 
if the heaviest neutrino has a mass larger than $\sim 0.1$ eV. 
Therefore, if the Z-burst scenario turns out to be correct,
it is in principle possible to measure a neutrino mass in this range
\cite{Pas:2001nd,Fodor:2001qy,Ringwald:2001mx}.

\subsection{Neutrino relativistic energy}

In addition to affecting structure formation neutrinos also 
contribute relativistic energy density in the early universe.
This has a profound effect on big bang nucleosynthesis, as well
as the CMBR formation.
The Friedmann equation $H^2 = 8 \pi G \rho$ yields the relationship
between temperature and expansion rate.
The beta reactions which keep the equilibrium between neutrons
and protons in the early universe freeze out roughly when
$\Gamma/H \sim 1$. If relativistic energy density is added then
$H$ is larger for a given temperature and the beta reactions 
freeze out faster. The neutron to proton ratio is in equilibrium
given by $n/p \propto e^{-Q/T}$, with $Q = m_n - m_p = 1.293$ MeV,
so that if relativistic energy
density is added more neutrons survive. This in turn means that
more helium is formed, and indeed observations of the
primordial helium abundance can be used to 
constrain the amount of relativistic energy density.

The standard way to parameterise such energy density is
in equivalent number of neutrino species $N_\nu \equiv \rho_R/\rho_{\nu,0}$,
where $\rho_{\nu,0}$ is the energy density of a standard neutrino
species. At present the bound is roughly $N_\nu \leq 4$
\cite{Lisi:1999ng}.
At first sight this bound can be translated directly into a bound on
the neutrino lepton number as well, because a non-zero lepton number
yields additional energy density.
\begin{equation}
N_\nu = 3 + \frac{30}{7}\left(\frac{\mu}{\pi T}\right)^2
+  \frac{15}{7}\left(\frac{\mu}{\pi T}\right)^4,
\end{equation}
assuming that only one neutrino species has non-zero lepton number.

However, there is a fundamental difference in that the bound is
flavour sensitive. The electron flavour neutrinos enter directly
into the beta reactions, and an electron neutrino lepton number
therefore has a different influence on BBN than muon or tau
neutrino lepton numbers \cite{Kang:xa}.

In practise this means that a large positive chemical potential
in muon and tau neutrinos can be compensated by a small electron
neutrino chemical potential.
Of course such models are quite contrived, but it is highly
desirable with independent methods for determining the cosmological
neutrino lepton numbers.

It turns out that the CMBR is at least in principle also an 
excellent probe of the relativistic energy density. The reason is
that an increase in the relativistic energy density delays matter
radiation equality, which in turn leads to an increase of the
so-called early integrated Sachs-Wolfe (ISW) effect. In the power
spectrum this shows up as an increase in power around the scale
of the particle horizon slightly after recombination, i.e\ around 
the scale of the first acoustic peak.

This effect has been used previously to constrain $N_\nu$ using
data from the best present day experiments 
\cite{Hannestad:2000hc,Hannestad:2001hn,Esposito:2000sv,Kneller:2001cd}, 
namely Boomerang \cite{boom},
Maxima \cite{max}, CBI \cite{cbi} and DASI \cite{dasi}. 
Unfortunately the data is not yet of sufficient
accuracy to yield constraints anywhere near as strong as BBN. However,
on the other hand they do not suffer from the same problems of being
flavour sensitive. It is therefore possible to combine BBN
and CMBR constraints to yield a non-trivial bound on neutrino
lepton chemical potentials \cite{Hansen:2001hi}. 
Interestingly, if neutrinos oscillate these
bounds can be very significantly strengthened \cite{Pastor:2001iu}.

In the near future a much more accurate determination of $N_\nu$ from
CMBR measurements will become possible thanks to the satellites
MAP and Planck. It was estimated by Lopez et al. 
\cite{Lopez:1999aq} that it would
be possible to measure $\Delta N_\nu \sim 0.04$ using Planck data.
However, this is probably overly optimistic and a more reasonable
estimate seems to be $\Delta N_\nu \sim 0.1-0.2$ \cite{Bowen:2001in}.

This will also allow for a possible detection of sterile neutrinos
mixing with ordinary neutrinos in the early universe over a wide
range of parameter space \cite{Hannestad:1998zg}.

%%%%%%%%%%%%%%%%%%%%%%%%%%%%%%%%%%%%%%%%%%%%%%%%%%%%%%%%%%%%%%%%%%%%%%%%
% Section IV
%%%%%%%%%%%%%%%%%%%%%%%%%%%%%%%%%%%%%%%%%%%%%%%%%%%%%%%%%%%%%%%%%%%%%%%%

\section{Supernova neutrinos}

Supernovae are, as far as we know, the most powerful neutrino
sources in the present day universe. In the timespan of a few seconds
a typical core collapse supernova emits about $10^{57}$ neutrinos
with an average energy of 10-15 MeV, or a total energy output of 
$\sim {\rm few} \times 10^{53}$ erg.
For this reason, supernovae are powerful laboratories for neutrino
physics and can be used to search for non-standard neutrino physics,
such as sterile neutrinos or neutrino magnetic moments.

However, the fundamental problem with supernovae are that within
our own galaxy they occur quite rarely on human timescales.
In order to detect neutrinos from a supernova it must be within
our own galaxy or the immediate vicinity. Even M31 is sufficiently
far away that only a few neutrinos would be detected by present 
day detectors.
So far the only detection of neutrinos from a supernova are
the 24 events from SN1987A detected by Kamiokande (11) \cite{Hirata:ad}, 
IMB (8) \cite{Bratton:ww} and Baksan
(5) \cite{baksan}.
The number and energy of these events are compatible with theoretical
expectations, except that the average energy of the events is
slightly lower than expected.

The fact that the duration of the observed neutrino signal matches
the expectations has been used to put an upper bound on the neutrino
mass. Indeed, if the emitted neutrinos have mass, their velocity
would be energy dependent and the signal would be more widely
spread in time. The time delay can be written as
\begin{equation}
\Delta t = 2.6 {\rm s} \,\, d_{50 \, {\rm kpc}} E_{10 \, {\rm MeV}}^{-2}
m_{\nu, 10 \, {\rm eV}}^2,
\end{equation}
and the observations therefore give roughly the bound $m_\nu \leq 20$ eV
\cite{Loredo:mk,Loredo:2001rx}.

Unfortunately this is much weaker than the bound from both tritium
decay experiments and from large scale structure observations.
Moreover, the prospects for increasing the sensitivity using a future
galactic supernova and the present day detectors is not too promising.
The only signal which would be sufficiently well located in time
would be the neutronisation signal of $\nu_e$'s. However, the detection
rate in water Cerenkov detectors is so low that even Super Kamiokande 
would only see about one event from a galactic supernova.
The reason is both that $\nu_e$ is detected via the elastic scattering
process $\nu_e e \to \nu_e e$, whereas the $\bar\nu_e$'s 
are detected via the charged current reaction $\bar\nu_e p \to ne^+$,
which has a much higher cross section, but also that the overall
flux of these neutrinos is much lower.
Therefore it would still be necessary to use the cooling phase
neutrinos, and the only real improvement would be that it would
be possible to measure the rise time of the cooling signal which
should be determined to an accuracy of about one second.
This should make a determination within about 2-3 eV possible.

Another possibility is if the proto neutron star collapses to a black
hole within a few seconds. This will lead to an immediate decrease
in neutrino luminosity, and if such a signature is observed from
a future galactic supernova, a mass determination in the eV
regime is possible \cite{Beacom:2000ng}.

Somewhat more promising than this is the possibility of measuring
neutrino oscillations from the observation of a galactic supernova.
A detector like SNO is able to measure different neutrino flavours.
Since neutrinos of different flavours are emitted with somewhat
different energies a flavour swap due to oscillations could 
be detectable. However, this still requires a sufficiently 
good theoretical understanding of the neutrino emission process.
At present, no full scale neutrino spectrum calculation including
self-consistently all the relevant microphysics has been performed.
Early calculations indicated that neutrinos of different flavour
are emitted with different energy. The reason is that their 
interaction cross sections are different, and that they are therefore
emitted from different regions of the proto-neutron star.
The average energy of emitted neutrinos is roughly given by
\cite{janka}
\begin{equation}
\langle E_\nu \rangle \sim \cases{
10-12 \, {\rm MeV} & for $\nu_e$ \cr
14-17 \, {\rm MeV} & for $\bar\nu_e$ \cr
24-27 \, {\rm MeV} & for $\nu_\mu,\nu_\tau,\bar\nu_\mu,\bar\nu_\tau$}
\end{equation}
However, these calculations did not include the effect of 
nucleon-nucleon inelastic scattering processes which for muon and tau
neutrinos are the most important thermalization processes
\cite{Raffelt:2001kv,Hannestad:1997gc,suzuki}.
It remains to be seen what influence the inclusion of such processes
will have on the neutrino spectra.

Dighe and Smirnov \cite{Dighe:1999bi}
have studied in detail the prospects for measuring
neutrino oscillations by observing the neutrino spectrum from
a future galactic supernova. At present the conclusion about
the possibility for measuring the various mixing parameters is
somewhat uncertain, mainly because of uncertainty in how well
a theoretical calculation of the neutrino emission can be performed.

%%%%%%%%%%%%%%%%%%%%%%%%%%%%%%%%%%%%%%%%%%%%%%%%%%%%%%%%%%%%%%%%%%%%%%%%
% Section V
%%%%%%%%%%%%%%%%%%%%%%%%%%%%%%%%%%%%%%%%%%%%%%%%%%%%%%%%%%%%%%%%%%%%%%%%

\section{Discussion}

Neutrinos are very weakly interacting particles and therefore 
parameters of the neutrino sector are difficult to measure.
So far the best information on the neutrino mixing matrix comes from
observations of neutrinos from the Sun or from cosmic rays.
Likewise the strongest upper bound on the absolute scale of neutrino
masses comes from observations of cosmological large scale structure.

In the coming years more new neutrino accelerator experiments will
come online and measure some of the neutrino parameters much 
more accurately. 
The coming years will likely see a shift in the 
emphasis of the relation between neutrinos and astrophysics. 
Neutrinos will no longer be a free parameter in astrophysical models
which can be invoked to explain puzzles. This in turn should allow
for a deeper understanding of such phenomena as supernovae.

However, some neutrino parameters will also in the future only be
accessible via astrophysical observations, and neutrino astrophysics
is likely to remain a very active field for years to come.

\Acknowledgements
I wish to thank Francesco Vissani, Vito Antonelli and Andreas Ringwald
for comments on the manuscript.

\end{document}